\documentclass[prl,aps,preprint]{revtex4-1}
\usepackage{graphicx}
\usepackage{amsfonts}
\usepackage{amssymb}
\usepackage{amsmath}
\usepackage{textcomp}

\usepackage{color}
\usepackage{psfrag}

\usepackage{epsfig}

\begin{document}

\title{Hyperuniform long-range correlations are a signature of disordered jammed hard-particle packings}

\author{Chase E.~Zachary,$^{1}$ Yang Jiao,$^{2}$ and Salvatore Torquato$^{1,2,3,4}$}
\email{torquato@princeton.edu}
\affiliation{$^{1}$Department of Chemistry, $^{2}$Department of Mechanical and Aerospace Engineering, 
$^{3}$Department of Physics, 
$^{4}$Princeton Center for Theoretical Science,
Princeton University, Princeton, New Jersey 08544, USA}

\begin{abstract}

We show that quasi-long-range (QLR) pair correlations that decay asymptotically with scaling $r^{-(d+1)}$ in $d$-dimensional Euclidean space $\mathbb{R}^d$, trademarks of certain quantum  systems and cosmological structures, are a universal signature of maximally random jammed (MRJ) hard-particle packings.  We introduce a novel hyperuniformity  descriptor in MRJ packings by studying local-volume-fraction fluctuations and show that infinite-wavelength fluctuations vanish even for packings with size- and shape-distributions.  Special void statistics induce hyperuniformity and QLR pair correlations.  

%We present strong evidence that quasi-long-range (QLR) pair correlations, trademarks of certain quantum many-body systems and cosmological structures, are a universal signature of maximally
%random jammed (MRJ) hard-particle packings.
%We utilize a novel descriptor of hyperuniformity in MRJ packings by examining local-volume-fraction fluctuations and show that infinite-wavelength fluctuations vanish even for packings with size- and shape-distributions.  
%Although the void spaces of MRJ packings are highly constrained by the underlying contact network, sufficient
%fluctuations in the sizes of the interparticle voids induce hyperuniformity and QLR pair correlations that decay asymptotically with scaling 
%$r^{-(d+1)}$ in $d$ Euclidean space dimensions, providing for the first time a concrete relationship between unexpected QLR correlations and local structure.

%Using a generalization of a conjecture due to Torquato and 
%Stillinger [Phys. Rev. E \textbf{68}, 041113 (2003)] that  all strictly jammed, saturated packings of hard particles
%lack infinite-wavelength local-volume-fraction fluctuations,  we show that the void spaces of MRJ packings
%are highly constrained by the underlying contact network. 

\end{abstract}

\maketitle

Quasi-long-range (QLR) pair correlations with asymptotic scaling $r^{-(d+1)}$ are unique trademarks of noninteracting spin-polarized fermionic ground states \cite{ToScZa08, ScZaTo09}, 
ground-state liquid helium \cite{ReCh67}, and the Harrison-Zeldovich spectrum 
of the early  Universe \cite{GaJoLa02}.  Such correlations also arise in maximally random jammed (MRJ) packings of monodisperse spheres \cite{DoStTo05}, which are prototypical glasses
\cite{FNChaikin} possessing maximal disorder among all jammed packings with diverging elastic moduli \cite{ToTrDe00, ToSt07, ToSt03}.  These packings were once thought to
describe the  structure of liquids \cite{Be65}, which typically possess pair correlations decaying exponentially fast \cite{ChLu00, ToSt03, ZaTo09, fn1}.
It is thus unclear why special QLR correlations occur in MRJ packings
and whether this behavior extends to packings of particles
with size- and shape-distributions, which possess nontrivial material properties  
\cite{torquato2002rhm, To86}. 

Noting that MRJ packings are structurally rigid with a well-defined contact network, Torquato and 
Stillinger conjectured \cite{ToSt03} that all strictly jammed, i.e., globally incompressible and nonshearable, saturated
packings of monodisperse spheres in $d$-dimensional real space $\mathbb{R}^d$ are hyperuniform, i.e.,
infinite-wavelength local-number-density fluctuations vanish \cite{ToSt03}, a proposition for which 
no counterexample has been found to date.  Although this conjecture made no mention of QLR pair correlations,
further work on 3D MRJ monodisperse sphere packings \cite{DoStTo05} established 
that
the structure factor $S(k)$
approaches zero linearly as the wavenumber $k\rightarrow 0$, inducing a QLR power-law tail $r^{-4}$ in
the pair correlation function $g_2(r)$, which is proportional to the probability density of finding a particle within a radial distance $r$ from a reference particle.  
However, neither study provided a quantitative explanation for the presence of QLR correlations in these systems, and the analysis does not blindly extend to MRJ packings of nonspherical and/or polydisperse particles in which the shape- and size-distributions of the particles are crucial.  
%However, it has not been clear how to generalize this conjecture to 
%MRJ packings of polydisperse and/or nonspherical particles; this problem requires choosing an appropriate descriptor of the packings that characterizes the QLR correlations of the system while incorporating nontrivial information about the shape and size distributions.  
Also, it is not obvious why MRJ packings are hyperuniform
since they lack the crystalline long-range order and Bragg peaks of periodic structures \cite{CoKu07}
and yet exhibit QLR pair correlations.
These unusual characteristics have important implications for understanding
the structural properties of MRJ and glassy material systems \cite{YaSc08,Za10}.   
We note that hyperuniform disordered point patterns 
have been used to create new materials with unusual optical properties \cite{BaStTo08} and large, complete
photonic band gaps \cite{FlToSt09b}.

In this Letter, we characterize universal properties of polydisperse and/or nonspherical MRJ hard-particle packings.
Specifically, we provide strong support for the claim that all saturated, strictly jammed particle packings,
which may be polydisperse and/or nonspherical, possess vanishing infinite-wavelength local-volume-fraction fluctuations, 
a recently-introduced extended notion of hyperuniformity \cite{ZaTo09}.
We thus provide the appropriate novel framework and descriptor to study universal QLR pair 
correlations in general MRJ packings.  
By examining binary circular disk, ellipse, and superdisk packings in $\mathbb{R}^2$, we show that infinite-wavelength density fluctuations 
associated with the particle centers do not vanish due to 
local microstructural inhomogeneities.  However, infinite-wavelength
local-volume-fraction fluctuations do vanish, so that the appropriate structural descriptor of MRJ packings is the two-point probability function $S_2(r),$ defined as the probability of finding both of two arbitrary points either within particles or in the exterior void space.  Importantly, our methodology contains all previously-reported results for monodisperse MRJ sphere packings  \cite{ToSt03, DoStTo05} as a special case, meaning that we provide a completely general means of understanding QLR correlations in jammed hard-particle packings.    
We show that the rigidity of the contact network places strong constraints on the size- and spatial-distributions
of the local voids, leading to hyperuniform QLR pair correlations that compete with the maximal randomness of the packings.  This observation supports the important notion that the 
topology and geometry of the void space \cite{fn4} are more basic than 
the particle space \cite{torquato2002rhm}.

It is sufficient for our purposes to consider 2D packings of binary circular disks, ellipses, and superdisks, 
but our arguments concerning the void space are general enough to 
incorporate higher-dimensional packings.
We use the Donev-Torquato-Stillinger  algorithm \cite{DoToSt05} to generate jammed packings of 
hard particles in $\mathbb{R}^2$.  Particles of two different sizes undergo event-driven molecular dynamics with periodic 
boundary conditions while simultaneously growing at a specified rate and fixed size ratio,
resulting in a final configuration that is essentially strictly jammed 
with a packing fraction $\phi \approx 0.8475$.  
The configurations that we study contain a small concentration ($\sim2.5\%$) of ``rattler''
particles, which are free to move within some small cage.  Saturation of the packing places an upper bound on the cage size, and the 
 rattler locations are roughly uniformly distributed in the simulation box.  Removing these 
rattlers breaks hyperuniformity by lifting the saturation constraint \cite{DoStTo05}; these particles are thus kept in our configurations,
which, by numerical protocol design, are close approximants to the MRJ state \cite{DoStTo05}.  
We have chosen particle concentrations $\gamma_{\text{small}} = 0.75$ and $\gamma_{\text{large}} = 0.25$ with size ratio $\beta = 1.4$ and have found that varying these parameters does not affect hyperuniformity. 

Hyperuniformity is determined by pair separation distances within either a point pattern or a heterogeneous medium.  Analysis of point patterns thus requires the pair correlation function $g_2$ or the associated structure factor $S(k) = 1+\rho\mathfrak{F}\{g_2(r)-1\}(k)$ with $\mathfrak{F}$ the Fourier transform.
%For point patterns, the \emph{pair correlation function}
%$g_2(r)$, proportional to the probability density of finding a particle within a radial distance $r$ from a reference particle, 
%is the function of interest.  The Fourier counterpart of $g_2(r)$ is the \emph{structure factor}
%$S(k) = 1+\rho \mathfrak{F}\{g_2(r)-1\}(k)$, where $\mathfrak{F}$ denotes the Fourier transform and $\rho$ is the number density.  
Using these functions, one can write the 
asymptotic behavior of the number variance $\sigma^2_N(R)$ in $\mathbb{R}^2$ as \cite{ToSt03}:
\begin{equation}\label{one}
\sigma^2_N(R) = 4\phi S(0) (R/D)^2 + \text{lower order terms},
\end{equation}
where $D$ is an effective length scale, taken here to be the average diameter ($D = \gamma_{\text{small}} D_{\text{small}} + \gamma_{\text{large}} D_{\text{large}}$)
of the particles, and $R$ is the radius of a circular observation window.  It follows from \eqref{one} that any point pattern with $S(0) = 0$ does not possess infinite-wavelength local number density fluctuations and thus is hyperuniform.  For heterogeneous media consisting of finite-volume inclusions, such as the binary MRJ
packings we study here, one should consider  the \emph{two-point probability function} $S_2(r)$, defined previously,
and the associated \emph{spectral density}
$\hat{\chi}(k) = \mathfrak{F}\{S_2(r) - \phi^2\}(k)$.  
Using these functions, the variance $\sigma^2_{\tau}(R)$ in the \emph{local volume fraction} $\tau$,
the volume fraction of a reference phase in an observation region, can be asymptotically written in $\mathbb{R}^2$ as \cite{ZaTo09}:
\begin{equation}\label{two}
\sigma^2_{\tau}(R) = [\rho/(4\phi)] \hat{\chi}(0) (D/R)^2 + \text{lower order terms}.
\end{equation}
As for point patterns, any heterogeneous medium with $\hat{\chi}(0) = 0$ is hyperuniform and lacks infinite-wavelength local-volume-fraction fluctuations.  Differentiating between the number variance and local-volume-fraction fluctuations is essential to characterize MRJ packings with a size distribution, and such a distinction has not previously been made in the literature. 

Both the structure factor and spectral density can be numerically obtained by direct Fourier transform \cite{BaStTo08} according to 
$S(\mathbf{k}) = N^{-1} \lvert \sum_{j=1}^N \exp(i \mathbf{k}\cdot \mathbf{r}_j)\rvert^2$ and
$\hat{\chi}(\mathbf{k}) = V^{-1} \lvert \sum_{j=1}^N \exp(i \mathbf{k}\cdot \mathbf{r}_j) \hat{m}(k; R_j)\rvert^2$,
where $\hat{m}(k; R) = \mathfrak{F}\{\Theta(R-\lVert\mathbf{r}\rVert)\}(k)$ is the Fourier transform of the indicator 
function for a disk of radius $R$, $\Theta(x)$ is the Heaviside step function, and $\left\{\mathbf{r}_j\right\}$ denotes 
the particle locations.  This result can be generalized 
to include particles of any geometry by changing the form of $\hat{m}$; the important restriction is that it 
will only apply for media consisting of impenetrable inclusions.  The reciprocal vectors $\mathbf{k}$
are discretized according to the shape of the simulation box.  For a square simulation box of side length $L$, the 
wave vectors are given by $\mathbf{k} = (2\pi/L) \mathbf{n}$ for $\mathbf{n} \in \mathbb{Z}^2$.  To obtain spherically-symmetric 
forms of the structure factor and spectral density, we radially average over all wavevectors within some small spherical shell in reciprocal space.
We have considered system sizes up to $N = 1000000$ particles to probe effectively the small-wavenumber region with even higher resolution, due to dimensional scaling, than previous studies \cite{DoStTo05}. 

\begin{figure}[!tp]
\centering
\includegraphics[width=0.875\textwidth]{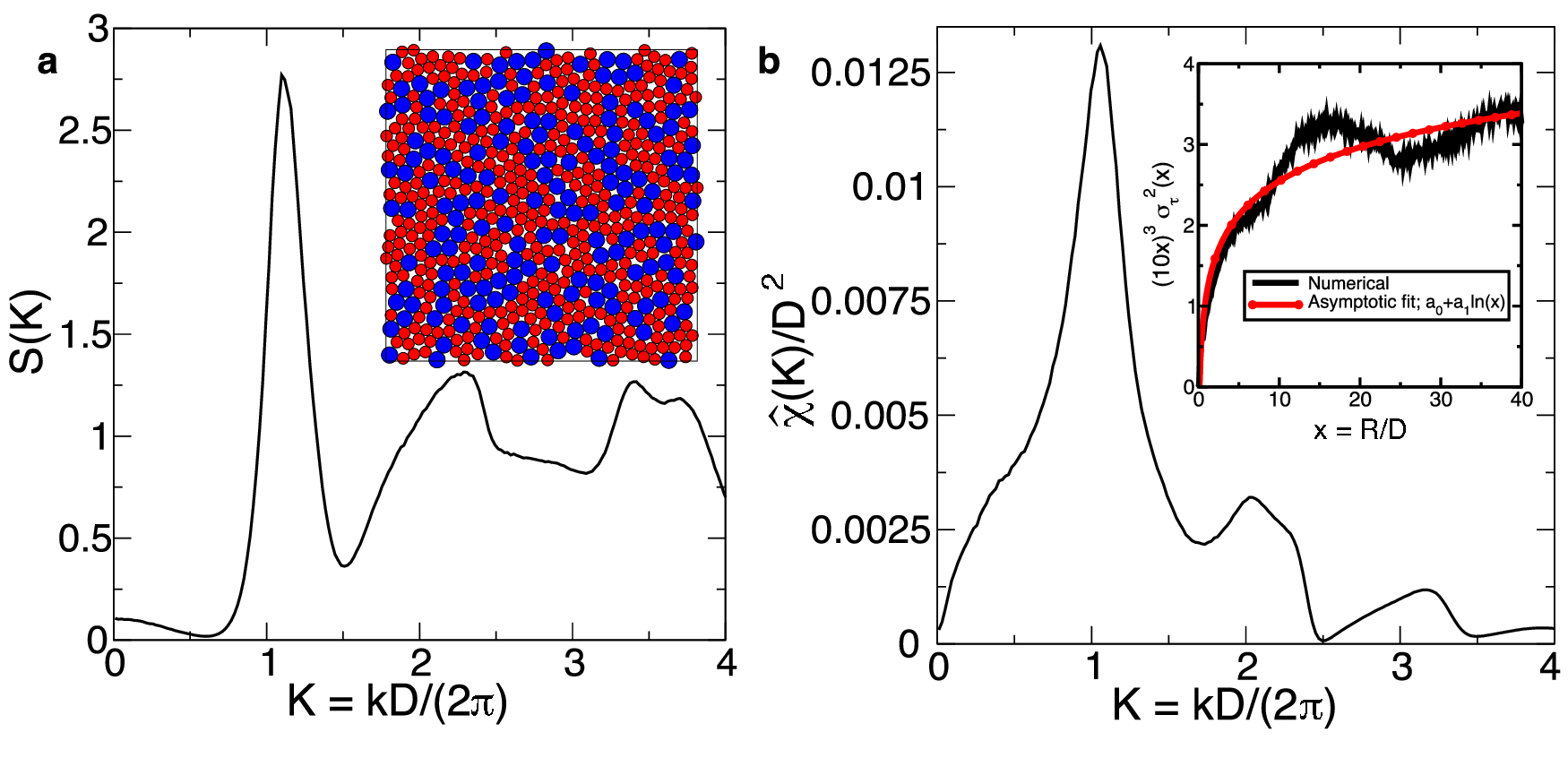}
\caption{(Color online)  (a) Structure factor and local configuration (inset) for the binary MRJ packing.  Large and small particles are shown in blue and 
red, respectively.  
(b) Corresponding spectral density with induced
local-volume-fraction fluctuations (inset).  The local-volume-fraction fluctuations decay logarithmically
faster than the volume of an observation window, implying both hyperuniformity and 
the presence of quasi-long-range correlations.}\label{figtwo}  
\end{figure}
Figure \ref{figtwo} depicts both the structure factor and spectral density of the binary MRJ 
disk packing.  The small-$k$ behavior of $S(k)$ is nonvanishing at the origin, showing that 
the point pattern generated by the disk centers is not hyperuniform, a result of the binary nature of the 
packing.  Figure \ref{figtwo} also shows that close-packed clusters of small particles permeate the system, but these local 
clusters are separated from each other by effective ``grain boundaries'' arising from the inclusion of large particles.
These grain boundaries preclude the suppression of 
infinite-wavelength local density fluctuations.
This result should be contrasted with previous calculations for 
MRJ packings of \emph{monodisperse} spheres in 3D \cite{DoStTo05}, where $S(k) \sim k$ 
as $k \rightarrow 0$.  It follows that the Torquato-Stillinger conjecture as originally 
stated for the number variance cannot hold for strictly jammed, saturated sphere  packings with a size distribution.  Nevertheless, 
the physical motivation behind this conjecture is intuitively appealing, and our results for the spectral density in Fig. \ref{figtwo}
show that it continues to hold in the aforementioned generalized sense upon analyzing local-volume-fraction fluctuations. 

We have fit the small-$k$ region of the spectral density with a minimal third-order polynomial of the form $\hat{\chi}(k) \sim \sum_{i=0}^3 a_i k^i$
and have found $a_0 = (1.0 \pm 0.2)\times 10^{-5}$, strongly 
indicating that this packing is hyperuniform.  
Thus, like 3D MRJ monodisperse sphere packings \cite{DoStTo05},
the spectral density is linear at the origin, implying that asymptotic local-volume-fraction fluctuations decay logarithmically faster than the
observation window volume, i.e., $\sigma^2_{\tau}(R) \sim (b_0 \ln(R) +b_1)/R^{3}$ as $R\rightarrow +\infty$.
%\begin{equation}
%\sigma^2_{\tau}(R) \sim (b_0 \ln(R) +b_1)/R^{3} \qquad (R\rightarrow +\infty).
%\end{equation}
This behavior implies that $S_2(r)$ contains a QLR power-law tail $r^{-3}$, extending to $r^{-(d+1)}$ in $\mathbb{R}^d$ \cite{ZaTo09}.  
%Fig. \ref{figtwo} also shows small-$k$ regions of the spectral densities for ``nearly jammed'' disk packings with $\phi < \phi_{\text{MRJ}}$.  Crucially,
%relaxing the jamming constraint destroys  hyperuniformity. 

\begin{figure}[!tp]
\centering
\includegraphics[width=0.65\textwidth]{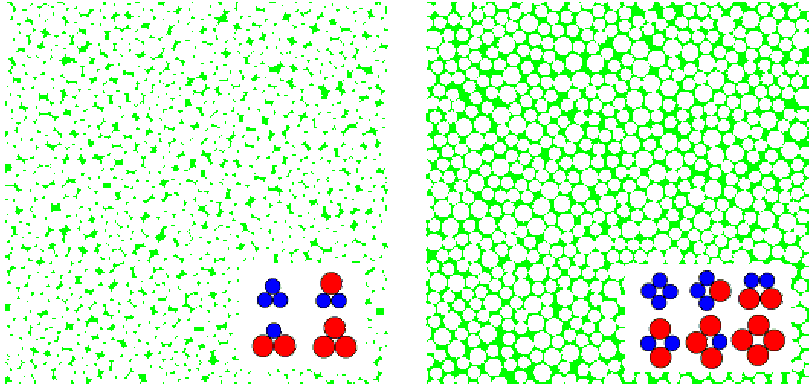}
\caption{(Color online) Jamming constraints on the void space.  The left panel shows the
MRJ state, which is dominated by three-particle loops of the types shown.  The right
panel is a nearly-jammed configuration with volume fraction $\phi = 0.75$, which 
exhibits an increasing fraction of skewed higher-order loops (also shown) as the void space 
becomes more connected.}\label{figthree}  
\end{figure}
To clarify the disparity between the number variance and local-volume-fraction fluctuations, we have examined the distribution of the void space.  
Since the spectral density measures 
both the inclusion phase and the 
surrounding matrix \cite{torquato2002rhm}, hyperuniformity is invariant to the choice of the reference phase, and homogeneity in 
the void phase is sufficient to induce hyperuniformity.  Figure \ref{figthree} highlights the void 
phase for the binary MRJ packing in addition to a nonhyperuniform  ``nearly jammed'' disk packing.  
The MRJ contact network is dominated by local three-particle contacts (or \emph{loops}), 
but by moving away from the jammed state, higher-order loops become increasingly common.  
This observation implies that the void-space distribution is inherently constrained by strict jamming.
%We have quantified the available void space in {Fig.~\ref{figthree}} using the \emph{cumulative pore-size distribution function} $F(\delta)$, which is the probability of 
%finding a spherical pore centered in the void space with radius $\delta$ less than or equal to the closest contact distance with the matrix-inclusion 
%interface \cite{torquato2002rhm}.
%Saturation and strict jamming of the MRJ packing enforce compact 
%support of the pore size distribution \cite{To10} and severely bound the pore sizes.
 It is true that for any strictly jammed, saturated 
packing of hard disks (2D) the matrix phase must be disconnected (see Fig. \ref{figthree}), which, although not essential for hyperuniformity,
implies that the void distribution is determined completely by the contacts
among the particles \cite{fn2}.  
In a companion paper \cite{ZaJiTo10}, we quantify the relationship between jamming and
the constrained void space by measuring the distribution of pore sizes and by utilizing rigorous bounds
that account for jamming and hyperuniformity.
%In a companion paper \cite{ZaJiTo10}, we will provide a direct connection between the jamming constraint and the shape/size distribution of the local voids.

Jamming is a critical factor for hyperuniformity to hold; even upon moving infinitesimally 
away from  jamming, hyperuniformity is broken.      
This behavior implies that knowledge of local $n$-particle contacts is equivalent to specifying the distribution of void sizes.
Relaxing the jamming constraint generates nonuniform, connected void-space regions that eventually percolate, skewing the distribution of pore sizes. 
Thus, the local void sizes  in the MRJ packing are inherently correlated with each other via the contact network. Specifically,
the presence of a large void limits the sizes of neighboring voids since the particles are strictly jammed.  The fact that hyperuniformity is lost
upon breaking the contact network implies that these correlations are inherently quasi-long-ranged, thereby suggesting a fundamental relationship 
between the distribution of the void space and spatially-extended correlations.

\begin{figure}[!tp]
\centering
\includegraphics[width=0.8\textwidth]{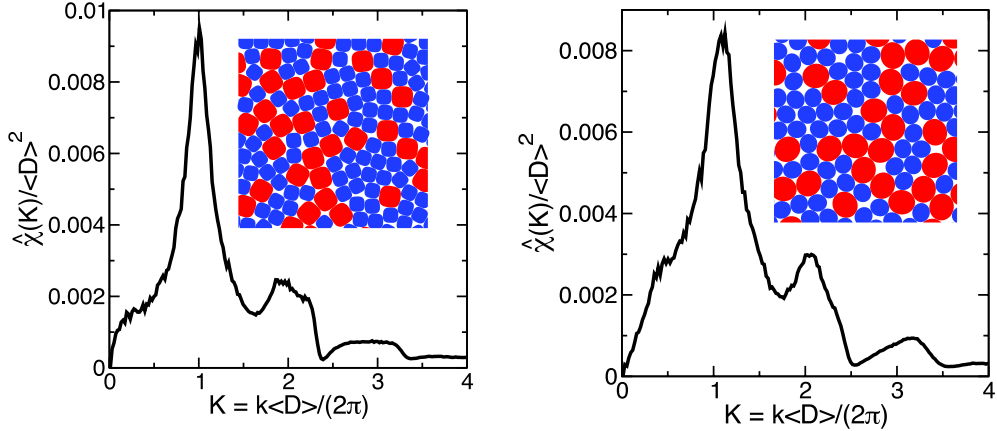}
\caption{(Color online)  Spectral densities for binary MRJ packings of (left) superdisks ($p = 1.5$) and (right) ellipses ($\alpha = b/a = 1.1$).  Also included
are portions of the jammed configurations (insets).  Large and small particles are in red and blue, respectively.}\label{figfour}  
\end{figure}
We have extended this analysis to MRJ packings of noncircular and nonspherical particles.  Figure \ref{figfour} shows the spectral densities associated with binary MRJ packings of hard 
superdisks and ellipses \cite{fn3}.
In both cases the packings are hyperuniform even though the inclusion shapes are no longer isotropic.  
Although the local contacts for these packings are less uniform than for hard disks, the 
constraints of saturation and strict jamming again limit the distribution of pore sizes and enforce hyperuniformity.
Also, the small-$k$ behaviors of 
the spectral densities for both the superdisk and ellipse packings are linear.

It has been previously observed that this small-wavenumber scaling is directly related to the ``degree of order'' of the packing \cite{GaJoTo08, DoStTo05}.  
Linear small-$k$ behavior in the spectral density therefore 
reflects a reasonable amount of variability in the shapes and sizes of the local voids.
Lower-order small-wavenumber terms in the spectral density (e.g., $\hat{\chi}(k) \sim k^{1/2}$) are associated with even greater
variability in the distribution of the void sizes and therefore reflect increasing probabilities of observing large voids.  These types of local-volume-fraction fluctuations are thus incompatible with 
strict jamming, and the small-wavenumber scaling of the spectral density is minimized to the smallest positive integer
value [$\hat{\chi}(k) \sim k$].
Our results thereby suggest that QLR correlations
in $S_2(r)$ as induced by a linear small-wavenumber region of the spectral density are a likely 
sufficient indicator of MRJ particle packings.  

We have established universal features of MRJ particle packings using local-volume-fraction fluctuations.
Our results generalize the QLR pair correlations of 3D MRJ sphere packings to saturated, strictly 
jammed hard-particle packings and provide insight into the origin of these
correlations.  
These packings are characterized by QLR pair correlations ($r^{-(d+1)}$ in $\mathbb{R}^d$) 
with a linear small-wavenumber region in the spectral density, and moving even slightly off the jamming point destroys hyperuniformity. 
We have justified this property using the variability of the
void space, which is both constrained by the local contact network and maximally disordered among all related saturated, strictly jammed packings.  
Our arguments incorporate higher-dimensional packings, and future work will 
examine MRJ packings of ellipsoids \cite{DoSiSaVa04} and superballs \cite{JiStTo10}.
We have also shown that local-number-density fluctuations of the particle centers are in general
 not sufficient to 
characterize MRJ packings because the inclusion shapes and sizes are rigorously linked to the void space,
meaning that the appropriate quantities to examine are the two-point probability function $S_2(r)$ and the associated spectral density $\hat{\chi}(k)$.
We mention that recent, independent work \cite{Be11} has examined fluctuation-response 
relations in MRJ packings of polydisperse spheres with similar conclusions to our own.  However, 
no mention is made of quasi-long-range correlations, and MRJ packings of nonspherical particles 
are not considered.
Our arguments provide insight into the structural properties of jammed particle packings and the nature of the jamming
transition  and suggest certain quantum  many-body systems and cosmological structures 
with special QLR correlations \cite{ToScZa08, ScZaTo09,ReCh67} are 
statistically ``rigid'' with vanishing infinite-wavelength number density fluctuations 
vanish and nonanalyticity in the structure factor at small wavenumbers.

\begin{acknowledgments}
This work was supported by the National Science Foundation under Grants DMS-0804431 and DMR-0820341.
\end{acknowledgments}


\begin{thebibliography}{26}
\bibitem{fn1}
Exponentially fast decay of pair correlations
implies that the structure factor $S(k)$ is analytic at the origin.
\bibitem{fn4}
The void space of a packing is exterior to the inclusions.
\bibitem{fn2}
For higher $d$, the particle and void space can both be connected,
but our arguments should hold since the contact network still constrains the local void geometries, inducing
correlations between void sizes.  Void-space percolation is thus not sufficient to break hyperuniformity.
\bibitem{fn3}
A superdisk is defined by the region
$\lvert x_1\rvert^{2p} + \lvert x_2\rvert^{2p} \leq 1$ for $p \geq 0$; similarly, an ellipse is given by $(\lvert x_1\rvert/a)^2 + (\lvert x_2\rvert/b)^2 \leq 1$
for $a, b > 0$.
\end{thebibliography}
\end{document}